\documentclass[useAMS,usenatbib]{mn2e}

\usepackage{amsmath} 
\usepackage{amssymb} 
\usepackage{graphicx} 
\usepackage{indentfirst} 
\usepackage{pdflscape} 
\usepackage{placeins} 
\usepackage{nicefrac} 
\usepackage{afterpage} 
\usepackage{rotating} 
\usepackage{booktabs} 
\usepackage{rotating} 
\usepackage{color} 
\newcommand{\kms}{${\rm km\,s}^{-1}$} 
 
\newcommand{\lya}{\mbox{${\rm Ly}\alpha$}}

\title[]{On the origin of excess cool gas in quasar host halos} 
\author[]{Sean D. Johnson$^{1}$\thanks{E-mail: seanjohnson@uchicago.edu}, Hsiao-Wen Chen$^{1}$, and John S. Mulchaey$^{2}$\\
$^{1}$Department of Astronomy \& Astrophysics and Kavli Institute for Cosmological Physics, The University of Chicago, Chicago, IL 60637, USA\\
$^{2}$The Observatories of the Carnegie Institution for Science, 813 Santa Barbara Street, Pasadena, CA 91101, USA}
\begin{document}

\date{\today}

\pagerange{
\pageref{firstpage}--
\pageref{lastpage}} \pubyear{2015}

\maketitle

\label{firstpage}
\begin{abstract}
	Previous observations of quasar host halos at $z\approx2$ have uncovered
	large quantities of cool gas that exceed what is found around inactive galaxies
	of both lower and higher masses.
	To better understand the source
	of this excess cool gas, we compiled an exhaustive sample of $195$ quasars
	at $z\approx1$ with constraints on chemically enriched, cool gas traced
	by Mg\,II absorption in background
	quasar spectra from the Sloan Digital Sky Survey.
	This quasar sample spans a broad range of luminosities
	from $L_{\rm bol}=10^{44.4}$ to $10^{46.8}\,{\rm erg\,s^{-1}}$
	and allows an investigation of whether halo gas properties
	are connected with quasar properties. 
	We find a strong correlation between luminosity and cool gas
	covering fraction.
	In particular, low-luminosity quasars exhibit a mean gas covering fraction
	comparable to inactive galaxies of similar masses,
	but more luminous quasars exhibit excess cool gas approaching
	what is reported previously at $z\approx2$.
	Moreover, $30-40\%$ of the Mg\,II absorption
	occurs at radial velocities of $|\Delta v|>300$ \kms\, from
	the quasar, inconsistent with gas
	bound to a typical quasar host halo.
	The large velocity
	offsets and observed luminosity
	dependence of the cool gas near quasars can be explained if the
	gas arises from:
	(1) neighboring halos correlated through large-scale structure at Mpc scales,
	(2) feedback from luminous quasars,
	or (3) debris from the mergers thought to trigger luminous quasars.
	The first of these scenarios is in tension with the lack of correlation
	between quasar luminosity and clustering while the latter two
	make distinct predictions that can be tested with
	additional observations.

\end{abstract}

\section{Introduction} \label{section:introduction}

In order to reproduce the low
stellar-to-halo mass ratios of high mass
galaxies \citep[e.g.][]{Conroy:2009, Behroozi:2013, Kravtsov:2014}
both semi-analytic \citep[e.g.][]{Benson:2003}
and hydrodynamic simulations of galaxy evolution
must incorporate strong feedback from active
galactic nuclei (AGN) and quasars
\citep[for a recent review, see][]{Kravtsov:2012}.
Moreover, hydrodynamic simulations of galaxy evolution
that incorporate AGN feedback are better able to reproduce
the properties of galaxies observed in emission
\citep[e.g.][]{Springel:2005, DiMatteo:2005, Sijacki:2007, Schaye:2010, Gabor:2012, Vogelsberger:2013, Li:2014}.
While promising, these implementations
of quasar feedback are subject to significant systematic
uncertainties driven by poorly constrained aspects of quasar and AGN
physics,
and direct observations of quasar feedback are available
for only a small number of systems \citep[for a recent review, see][]{Fabian:2012}.

The low-density gas of the circum-galactic medium (CGM) provides
a sensitive laboratory for discriminating between possible feedback
models, and different feedback prescriptions result in order-of-magnitude
changes in CGM observables predicted in hydrodynamic simulations
of galaxies \citep[][]{Hummels:2013, Ford:2013, Shen_S:2013, Agertz:2014, Suresh:2015}.
The density of the CGM is nearly always too low to be studied in emission
with existing facilities, but significant progress can be
made by studying the gas in absorption when bright background
objects are serendipitously found at low projected distances, $d$,
from foreground galaxies.

Over the last decade, large samples totaling
nearly one thousand galaxies at $z\approx0$ to $2$ with constraints on
extended gas properties from absorption spectroscopy
have been assembled using a combination
of ground and space-based telescopes.
In particular, observations of the
H\,I Lyman series
\citep[e.g.][]{Chen:1998, Tripp:1998, Wakker:2009, Stocke:2013, Rudie:2013, Tumlinson:2013}
and doublet transitions due to heavy element ions such as
Mg\,II \citep[e.g.][]{Bowen:1995, Chen:2010a, Gauthier:2010, Bordoloi:2011, Lovegrove:2011},
C\,IV \citep[e.g.][]{Chen:2001b, Borthakur:2013, Liang:2014, Bordoloi:2014},
and O\,VI \citep[e.g.][]{Chen:2009, Wakker:2009, Prochaska:2011, Tumlinson:2011, Mathes:2014, Stocke:2014, Turner:2014, Johnson:2015}
have been particularly fruitful.

Recently, observing campaigns leveraging large imaging
and spectroscopic surveys such as the
Sloan Digital Sky Survey \citep[SDSS;][]{York:2000}
and dedicated campaigns on large, ground-based
telescopes have extended the study of the relationship
between galaxy and halo gas properties to galaxies
hosting quasars using projected quasar-quasar pairs. 
Foreground quasars at both $z\approx1$ and $z>2$ exhibit
a high incidence of optically thick, metal-enriched
absorption systems traced by H\,I \lya, C\,II, C\,IV, and Mg\,II absorption
along the transverse direction
at $d\lesssim300$ kpc
but low incidence
along the foreground quasar sightline itself
\citep[][]{Bowen:2006, Hennawi:2006, Hennawi:2007, Farina:2013, Hennawi:2013, Prochaska:2013, Farina:2014,
Prochaska:2014}.
This contrast indicates that the ionizing emission
from quasars is highly anisotropic, in qualitative agreement
with the unified theory of AGN \citep[e.g.][]{Antonucci:1993, Netzer:2015}.

The high incidence of optically thick gas at $d<300$ kpc
from quasars at $z\approx2$ is 
in significant excess relative to that found
for inactive galaxies both at $z<1$ \citep[][]{Chen:2010a, Gauthier:2010, Lovegrove:2011}
and at $z\approx2$ \citep[][]{Rudie:2012}.
Insights into the possible origin of this excess gas can be gained
from state-of-the-art hydrodynamic simulations that
include stellar but not AGN feedback \citep[e.g.][]{Fumagalli:2014}.
In particular, simulations with stellar feedback are
able to comfortably reproduce the observed incidence of Lyman-limit
systems around inactive, Lyman-break galaxies at $z\approx2$
but under-predict the incidence around quasar hosts by more
than a factor of two \citep[][]{Faucher:2015}.
This discrepancy suggests that AGN feedback may
be responsible for the excess observed in absorption
around quasar hosts.
Alternatively, the discrepancy could be the result of
an inability of the simulations to resolve the physical
scales relevant to the formation or survival of
cool gas clouds in massive halos \citep[][]{Fumagalli:2014, Meiksin:2015}
or if a substantial portion of the gas arises
in less luminous galaxies neighboring the quasar hosts
\citep[][]{Rahmati:2015, Suresh:2015}.

To better understand the relationship between
AGN activity and halo gas, 
we searched
the SDSS Data Release 12 \citep[][]{Eisenstein:2011, Alam:2015}
and compiled an exhaustive sample
of $195$ quasars at $z\approx1$ with constraints
on Mg\,II absorption at $d<300$ kpc from background quasars.
This large dataset enables a search for correlations between quasar properties
and extended, cool circumgalactic gas around the quasar hosts.

The paper proceeds as follows: In Section \ref{section:sample} we
describe the quasar sample and corresponding absorption-line measurements.
In Section \ref{section:absorption}, we characterize the absorption
as a function of projected distance, redshift, and quasar luminosity.
In Section \ref{section:discussion}, we discuss the implications
of our findings.

Throughout the paper, we adopt a $\Lambda$-cosmology
with $\Omega_{\rm m}=0.3$, $\Omega_\Lambda = 0.7$,
and $H_0 = 70\,{\rm km\,s^{-1}\,Mpc^{-1}}$.
All magnitudes are in the $AB$ system \citep[][]{Oke:1983} and
corrected for foreground Milky Way extinction following
\cite{Schlafly:2011}.

\section{Quasar sample}
\label{section:sample}

\begin{table*}
	\caption{Summary of quasar and absorption properties. The full table is available in the on-line version of the paper.}
	\label{table:sample}
	\begin{tabular}{cccccccccc}
	\hline
	& \multicolumn{4}{c}{Foreground quasar} & & & \multicolumn{2}{c}{Absorption properties} \\  \cline{2-5} \cline{8-9}
        & Right Ascension & Declination & & & $\Delta \theta$ & $d$ & $W_{\rm r}(2796)^{\rm b}$ & $\Delta v$ \\
	Name & (J2000) & (J2000) & $z^{\rm a}$ & $\log\,L_{\rm bol}/{\rm erg\,s^{-1}}$ & ($''$) & (kpc) & (\AA) & (\kms) & Ref.$^{\rm c}$ \\
	\hline \hline
J$09$$54$$+37$$34$ & $09$:$54$:$54.70$ & $+37$:$34$:$19.7$ & $1.544$ & $46.3$ & $3.1$ & $26$ & $1.10\pm0.17$ & $660$& $1$ \\
J$08$$36$$+48$$41$ & $08$:$36$:$49.40$ & $+48$:$41$:$50.0$ & $0.657$ & $45.7$ & $4.1$ & $28$ & $1.90\pm0.11$ & $-130$& $1$ \\
J$08$$42$$+47$$33$ & $08$:$42$:$57.37$ & $+47$:$33$:$42.6$ & $1.556$ & $46.5$ & $3.4$ & $29$ & $3.70\pm0.30$ & $-790$& $4$ \\
J$15$$50$$+11$$20$ & $15$:$50$:$43.59$ & $+11$:$20$:$47.5$ & $0.4358$ & $46.0$ & $5.2$ & $29$ & $<0.25$ & n/a& $4$ \\
J$11$$06$$+46$$35$ & $11$:$06$:$17.17$ & $+46$:$35$:$24.5$ & $1.602$ & $46.5$ & $4.4$ & $37$ & $1.12\pm0.15$ & $-130$& $4$ \\
J$11$$08$$+33$$06$ & $11$:$08$:$07.90$ & $+33$:$06$:$11.3$ & $1.502$ & $46.5$ & $5.5$ & $46$ & $5.75\pm0.26$ & $810$& $4$ \\
J$09$$38$$+53$$17$ & $09$:$38$:$04.21$ & $+53$:$17$:$43.9$ & $2.063$ & $45.9$ & $5.7$ & $47$ & $0.48\pm0.04$ & $310$& $3$ \\
J$23$$12$$+14$$44$ & $23$:$12$:$52.70$ & $+14$:$44$:$58.6$ & $0.7678$ & $45.2$ & $6.4$ & $47$ & $0.39\pm0.12$ & $-100$& $1$ \\
J$14$$27$$-01$$21$ & $14$:$27$:$58.88$ & $-01$:$21$:$30.3$ & $2.281$ & $46.6$ & $6.2$ & $51$ & $0.45\pm0.02$ & $-80$& $3$ \\
J$09$$09$$+16$$29$ & $09$:$09$:$57.08$ & $+16$:$29$:$06.5$ & $0.7275$ & $45.2$ & $7.1$ & $51$ & $<0.15$ & n/a& $4$
 \\\hline \multicolumn{9}{l}{\bf Notes} \\
		\multicolumn{9}{l}{$^{\rm a}$ Quasars with redshifts from narrow [O\,II] or [O\,III]  are shown to four decimal places and three decimal places otherwise.} \\
		\multicolumn{9}{l}{$^{\rm b}$ For non-detections, we report $3$-$\sigma$ upper limits integrated over a $250$ \kms\, velocity interval.} \\
		\multicolumn{9}{l}{$^{\rm c}$ Reference: $1\rightarrow$ \protect \cite{Bowen:2006}, $2\rightarrow$ \cite{Farina:2013, Farina:2014}, $3\rightarrow$ \cite{Prochaska:2014}, $4\rightarrow$ This work} \\
	\end{tabular}
\end{table*}

To compile a large sample of foreground-background
quasar\footnote{Throughout this paper, we refer to objects with broad-line emission and power-law dominated continua as quasars irrespective of the sub-division of AGN into quasars and Seyferts based on luminosity.} pairs with constraints on Mg\,II absorption,
we retrieved a list of the $395,281$
quasars classified by the SDSS-III automated
classification and redshift measurement pipeline \citep[][]{Bolton:2012}
as of Data Release 12.
From this sample, we selected foreground-background quasar pairs
that meet the following criteria:
\begin{enumerate}
	\item The projected distance between the foreground and background quasars at the foreground quasar redshift satisfies $d<300$ kpc.
	\item The velocity difference between the foreground and background quasars satisfies $\Delta v(z_{\rm b}, z_{\rm f}) < -10,000$ \kms.
	\item The expected wavelengths of the Mg\,II doublet at the redshift of the foreground quasar is covered by the SDSS spectrum of the background quasar.
	\item The expected wavelengths of the Mg\,II doublet is outside of the \lya \,forest in the background quasar spectrum. and
	\item The signal-to-noise ratio in the background quasar spectrum is sufficient to detect a moderate strength Mg\,II absorption system with rest-frame equivalent width of $W_{\rm r}(2796)>0.3 $ \AA\,
at $3$-$\sigma$ significance
at the foreground quasar redshift (a signal-to-noise ratio of $20$ per SDSS resolution element).
\end{enumerate}
The upper limit on the projected distance was chosen to
correspond to the expected virial radii of
quasar host halos at $z>0.4$.
The requirement that $\Delta v < -10,000$ was chosen to
avoid confusion with gas outflowing from the background
quasar \citep[e.g.][]{Wild:2008}.
The requirement that the wavelengths of the Mg\,II doublet
at the foreground quasar redshift fall outside of the \lya\,
forest in the background quasar spectrum was chosen to ensure that absorption
attributed to Mg\,II is not the result of coincidental \lya\, absorption systems.
Finally, the requirement that the signal-to-noise ratio in the background quasar
spectrum is sufficient to detect absorption systems of
$W_{\rm r}(2796)>0.3 $ \AA\,
was chosen to correspond to the typical
sensitivities of existing studies of Mg\,II absorption
around quasars at $z\approx1$ \citep[e.g.][]{Farina:2013, Farina:2014}.

The search yielded a sample of $195$ foreground-background
quasar pairs which we visually inspected to ensure that the automated
classifications as broad-line (Type 1) quasars and redshifts
from the SDSS database are robust.
For each of the $195$ quasar pairs, we then measured foreground
quasar properties and Mg\,II absorption properties as described in
Sections \ref{section:quasar_measurements} and
\ref{section:absorption_measurements} respectively.
The sample is summarized in Table \ref{table:sample}.

\subsection{Foreground quasar redshifts and luminosities}
\label{section:quasar_measurements}
Quasar redshifts from the SDSS pipeline are biased by
$\approx 600$ \kms\, due to asymmetric, blue-shifted emission profiles
of broad-line region emitting gas
\citep[][]{Gaskell:1982, Tytler:1992, Richards:2002, Hewett:2010}.
This bias exceeds the velocity difference typically
found between galaxies and associated Mg\,II
absorption at $d<150$ kpc \citep[][]{Chen:2010a} so more accurate
redshifts are required. When available, we adopted
redshifts from \cite{Hewett:2010} and otherwise,
we calculated the quasar redshifts using the template
and cross-correlation techniques described in \cite{Hewett:2010}.
In addition, for quasars with narrow [O\,II] or [O\,III] emission,
we measured the redshifts from these narrow emission lines
by fitting gaussian profiles and adopting the rest-frame
line centroid wavelengths from \cite{Hewett:2010}.

Uncertainties in narrow-line
based redshifts due to centroid uncertainties
are typically small, $\sigma \approx 30$ \kms,
but [O\,III] redshifts can be blueshifted due to
outflows in the narrow-line region \citep[][]{Boroson:2005}.
To evaluate the significance of this bias,
we compared [O\,II] and [O\,III] emission redshifts
for quasars within our SDSS sample and found
a mean bias of $\Delta v (z_{\rm O\,II}, z_{\rm O\,III}) = -30$ \kms\,
with a $1$-$\sigma$ scatter of $70$ \kms.

Broad-line based cross-correlation redshifts exhibit larger scatter 
with significant, non-Gaussian wings
in the redshift error distribution due to population variance in broad-line profiles.
To evaluate the uncertainties in our broad-line based redshifts,
we remeasured the cross-correlation redshifts for quasars with narrow-line emission
but with the narrow-lines masked during the cross-correlation.
We then compared these broad-line redshifts with the narrow-line
redshifts. The bias in the broad-line based cross-correlation
redshifts is consistent with zero and the $68\%$, $95\%$, $99\%$, and $99.7\%$
uncertainties correspond to $150, 370, 1000$, and $1500$ \kms\,respectively.

In addition to improved redshifts, we measured quasar bolometric
luminosities based on monochromatic, continuum luminosity measurements
and the bolometric corrections from \cite{Richards:2006}.
The monochromatic luminosities were estimated by fitting a power-law
continuum plus Fe\,II template \citep{Vestergaard:2001} model to
line-free continuum regions of the SDSS quasar spectra with rest-frame wavelength intervals
of $\lambda_r = 1350-1360, 1445-1465, 1700-1705, 2155-2400, 2480-2675, $
and $2925-3500$ \AA.
The uncertainties in the quasar bolometric luminosities
are dominated by sample variance in quasar spectral energy
distributions which result in errors of $\approx0.3$ dex
\citep[see discussion in][]{Richards:2006}.

 \begin{figure*}
	\begin{tabular}{cc}
	\includegraphics[scale=0.42]{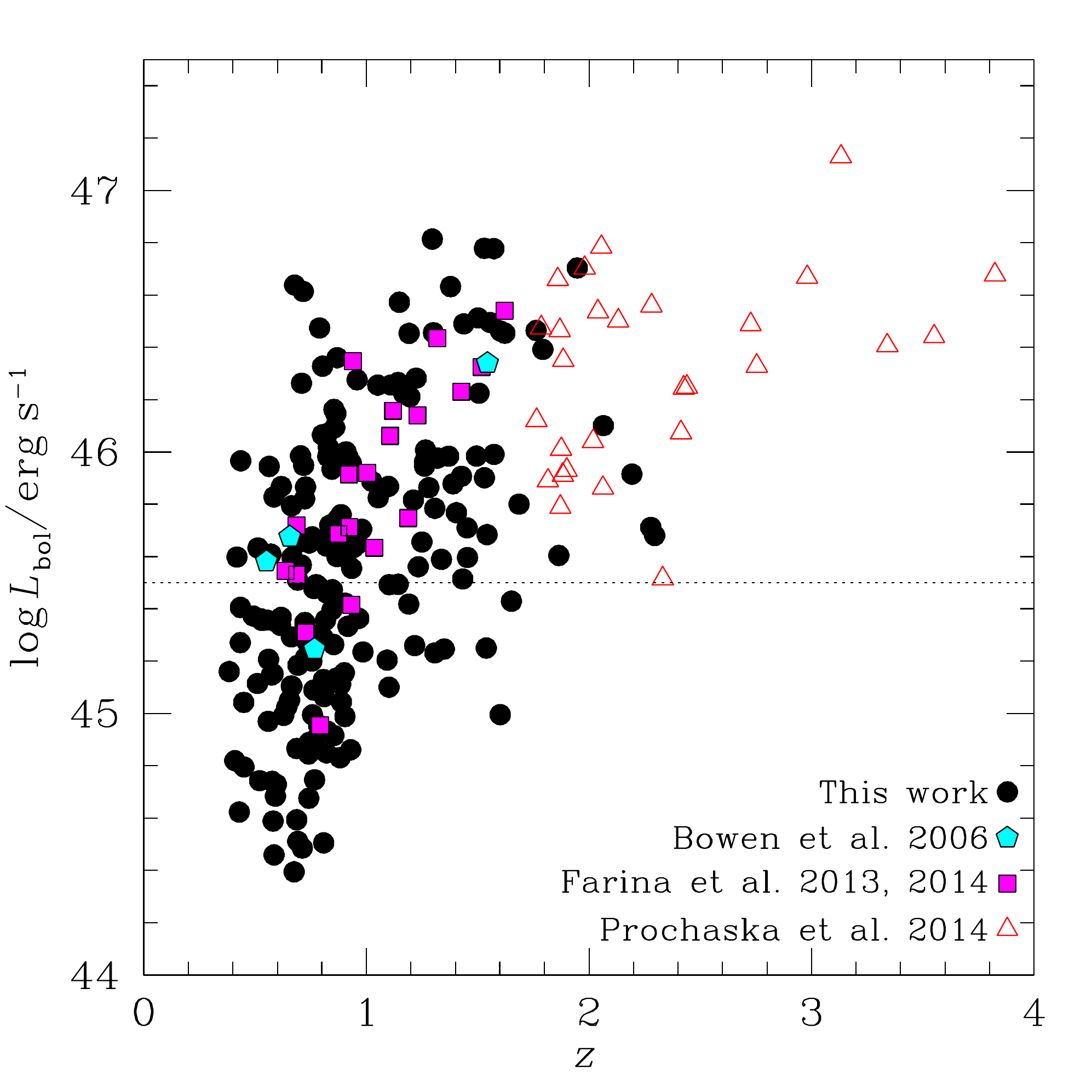} & \includegraphics[scale=0.42]{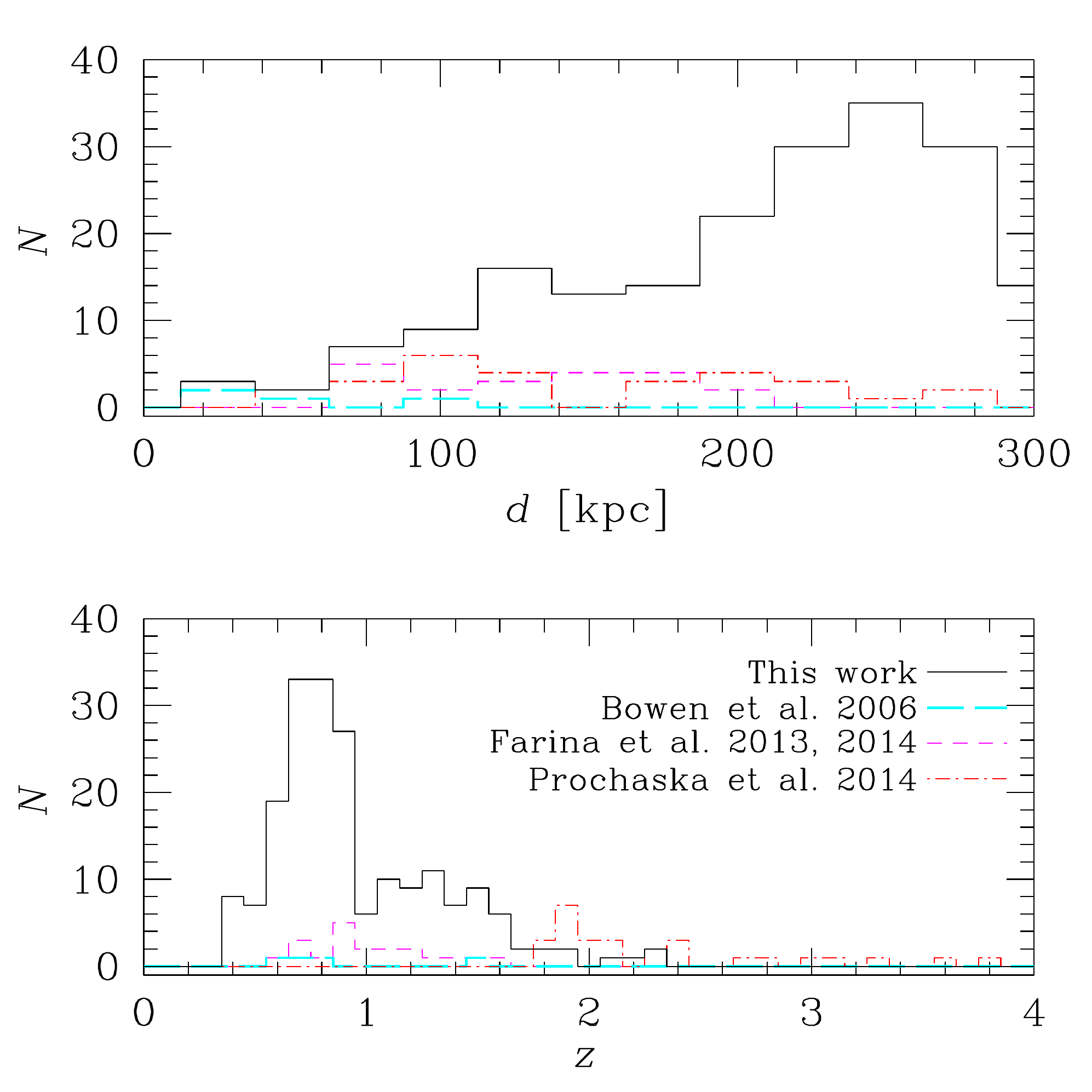}
	\end{tabular} 
	\caption{{\it left}: Bolometric luminosity of the foreground quasars
	               versus foreground quasar redshift. The horizontal
	               dotted line marks $\log\,L_{\rm bol}/{\rm erg\,s^{-1}}=45.5$,
	               the division between the luminous and low-luminosity
	               quasars used throughout the paper.
	              {\it right}: Projected distance ({\it top})
	              and redshift ({\it bottom}) histograms.
	              In all three panels, quasars from \protect \cite{Bowen:2006},
	              \protect \cite{Farina:2013} or \protect \cite{Farina:2014},
	              \protect \cite{Prochaska:2014}, and this work are displayed in
	              cyan, magenta, red, and black respectively, with symbols and
	              line-styles indicated in the legend.
	              The quasars from \protect \cite{Bowen:2006},
	              \protect \cite{Farina:2013}, 
	              \protect \cite{Farina:2014},
	              and this work have available constraints on the
	              transverse Mg\,II absorption from the
	              spectra of background quasars.
	              Mg\,II absorption constraints are unavailable
	              for the quasar sample from
	              \protect \cite{Prochaska:2014} due to the higher redshift range.
	              For these quasars,
	              C\,II rather than Mg\,II serves as a signature of cool, 
	              high H\,I column density gas.}
	\label{figure:sample}
\end{figure*}

\begin{figure*}
	\includegraphics[scale=0.5]{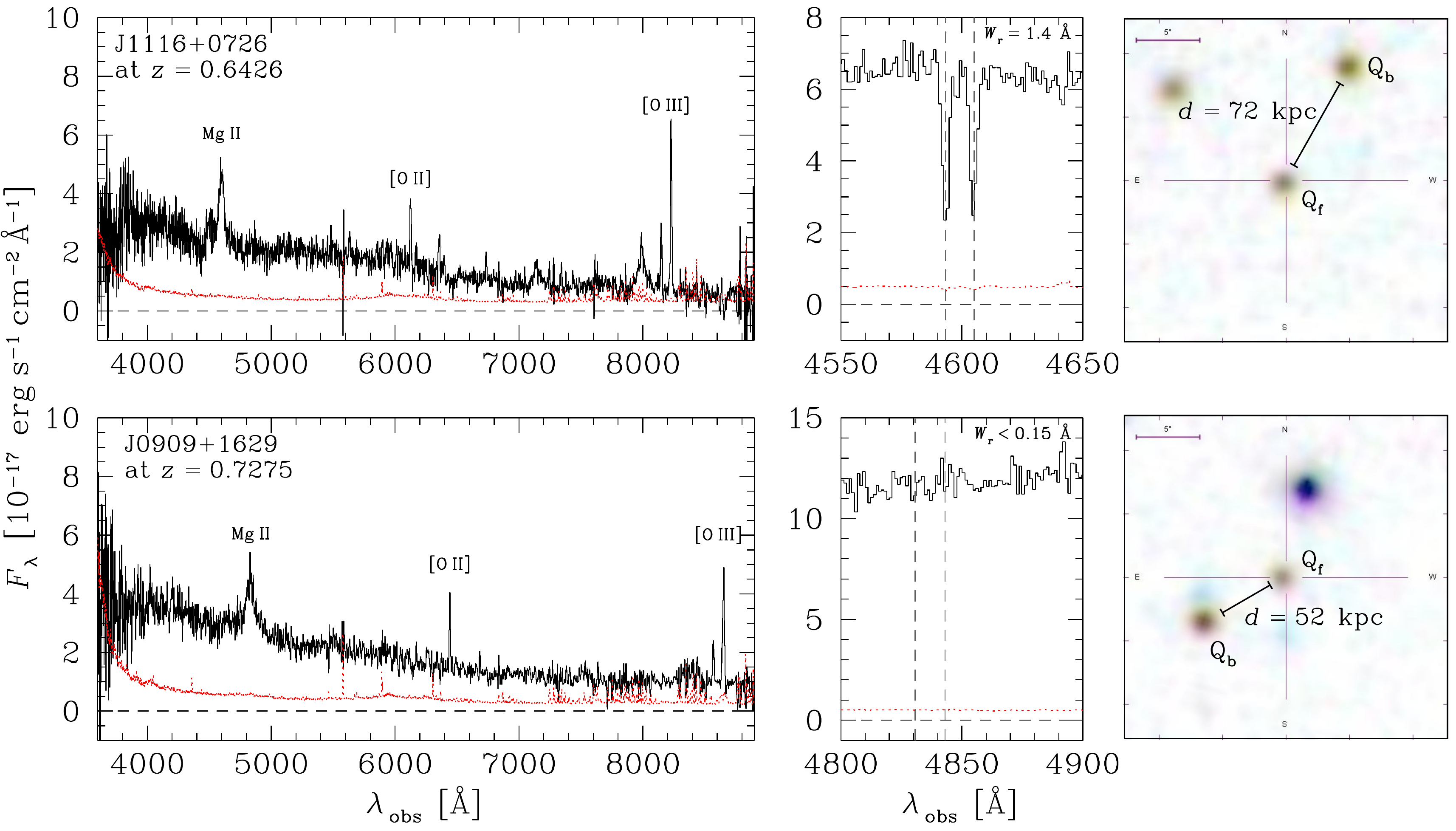}
	\caption{Two example quasars probed in Mg\,II absorption.
	               The {\it top} panels display a quasar
	              detected in Mg\,II absorption at $d=72$ kpc and the bottom panels display a quasar
	              without detected Mg\,II absorption at $d=52$ kpc.
	              The {\it left} panels display the SDSS spectra of the foreground quasars.
	              The {\it middle} panels display the SDSS spectra of the background quasars with
	              the expected positions of the Mg\,II doublet members at the systemic redshift of
	              the foreground quasar marked by vertical dashed lines. In both the left and middle
	              panels, the flux and error arrays are shown in black histogram and dotted red line respectively.
	              The zero-flux level is marked by a horizontal, black dashed line.
	              The SDSS image of the
	              fields around the quasars are shown on the {\it right} with the position of the foreground
	              and background quasars labelled as ${\rm Q_f}$ and ${\rm Q_b}$ respectively.
	              The purple
	              line at the top left of each image is $5''$ in length.}
	\label{figure:examples}
\end{figure*}

In order to include previously reported quasars with
absorption line constraints at $d<300$ kpc, we measured
bolometric luminosities for quasars from
\cite{Bowen:2006, Farina:2013, Farina:2014} and \cite{Prochaska:2014}
with available SDSS spectra using the same techniques.
For the quasars from \cite{Farina:2013} and \cite{Farina:2014}
without SDSS spectra
we adopted the bolometric luminosity estimates
reported by \cite{Farina:2013} and \cite{Farina:2014}
which were measured using a similar continuum fitting procedure.
For those quasars from \cite{Bowen:2006} and \cite{Prochaska:2014}
without public spectra, we estimated luminosities
by finding close matches in redshift and apparent
magnitude space in the quasar sample from \cite{Shen:2011}
which has monochromatic luminosities estimated from
continuum fitting. These broad-band based luminosities
reproduce spectroscopic
measurements with a standard deviation of $0.2$ dex.

The redshift, luminosity, and projected distance distributions of the resulting
quasar catalogs are displayed in the Figure \ref{figure:sample}.
The samples from \cite{Bowen:2006}, \cite{Farina:2013}, \cite{Farina:2014},
and this work
are characterized by mean redshifts of $\langle z \rangle \approx 1$
while the sample from \cite{Prochaska:2014} is characterized
by $\langle z \rangle = 2.2$.
The quasar sample presented in this paper
spans a luminosity range of
$\log\,L_{\rm bol}/{\rm erg\,s^{-1}}=44.4$ to $46.8$
(see the left panel of Figure \ref{figure:sample}),
extending the study of gas in absorption around
quasars to lower luminosities than were previously available.

\subsection{Absorption-line measurements}
\label{section:absorption_measurements}

In order to place constraints on the transverse Mg\,II
absorption near the foreground quasars, we visually
searched the background quasar spectra
for possible Mg\,II $\lambda 2796$ absorption within
$|\Delta v| < 1500$ \kms\, of the foreground quasar
redshift. The large search window was chosen to ensure
that uncertainties in broad-line based quasar
redshifts do not bias the results of this study.
In addition, this velocity search window is the same as
was used by  \cite{Prochaska:2014} which simplifies
the comparison between that work and our $z\approx 1$ sample.

When absorption was identified,
we measured the absorption equivalent
width and centroid in the following steps.
First, we locally fit the continuum by
defining feature-free continuum regions
at lower and higher wavelengths
and fit an outlier-resistant
line to these continuum regions.
We then continuum normalized the flux and error arrays
and calculated the rest-frame equivalent width and equivalent width
error by direct integration over a user-defined
interval.
We accepted the absorption according to a
$3$-$\sigma$ detection threshold and additionally
required that the identification of the absorption
feature as Mg\,II $\lambda 2796$ be confirmed
by the presence of at least one other absorption
feature at the same redshift
(e.g. Mg\,II $\lambda 2803$, Mg\,I $\lambda 2852$, 
or Fe\,II $\lambda \lambda 2344, 2374, 2382, 2586, 2600$).
Finally, we measured the absorber centroid
and full-width-at-half-maximum (FWHM)
by fitting a gaussian profile to the Mg\,II $\lambda 2796$
absorption and used the centroid to measure the
absorber redshift.
In a few cases, multiple absorption components
are found at $|\Delta v| < 1500$ \kms. For these
quasars, we report the total Mg\,II $\lambda 2796$
equivalent width and report $\Delta\,v$ for the
strongest component. Each of these cases is discussed
further in Section \ref{section:kinematics}.

In the case of non-detections, we placed $3$-$\sigma$ upper
limits on the Mg\,II $\lambda 2796$ equivalent width by
integrating the continuum normalized error array over a $250$
\kms\, window centered at the foreground quasar redshift.
The $250$ \kms\, window corresponds to the mean FWHM
of Mg\,II $\lambda 2796$ detections in our sample.
Two example quasars probed in absorption, one with detected Mg\,II absorption
and one without, are showcased in Figure \ref{figure:examples}
to demonstrate the data quality afforded by the SDSS spectra.

In one case, J$1250$$-$$0105$,
we identified possible Mg\,II $\lambda 2796$ absorption with 
$3$-$\sigma$ significance but were unable to confirm the identification
by detection of another line. For this quasar, we placed a limit
on the Mg\,II $\lambda 2796$ equivalent width by measuring the limit on 
Mg\,II $\lambda 2803$ and multiplied by two (the ratio
of the oscillator strengths of the two transitions).
We note that the results presented in this paper do
not change if the possible Mg\,II $\lambda 2796$
absorption for this quasar is treated as a detection.

In a few cases, we detected absorption
from the Mg\,II doublet but with the $\lambda 2796$
member contaminated by heavy element absorption
from another absorption system at a distinct redshift.
For these objects, we fitted the
Mg\,II and contaminating absorption profiles
using the VPFIT package \citep[][]{Carswell:1987, VPFIT}
and measured Mg\,II $\lambda 2796$ equivalent width
from the model fit.

Finally, we added absorption-line measurements from the
foreground-background quasar samples previously reported
in the literature in \cite{Bowen:2006, Farina:2013, Farina:2014}, and \cite{Prochaska:2014}.
The sample from \cite{Bowen:2006} contains four foreground quasars
probed at $d=26, 29, 47,$ and $98$ kpc all of which are detected
in Mg\,II absorption with equivalent widths
ranging from $W_{\rm r}(2796) = 0.4$ to $1.9$ \AA.
We included quasars from \cite{Farina:2013} and \cite{Farina:2014}
after converting their $2$-$\sigma$ detection limits to $3$-$\sigma$,
applying a uniform cut requiring sufficient signal in the background
quasar spectrum to detect an absorption system of $W_{\rm r}(2796) > 0.3$ \AA,
and restricting the Mg\,II search window to $|\Delta v| < 1500$ \kms.

The quasar sample from \cite{Prochaska:2014} spans a redshift
range of $z=1.8-3.8$, and consequently Mg\,II absorption constraints
are not available with the existing optical spectra.
Mg\,II absorption traces cool ($T\sim10^4$ K) gas
\citep[][]{Bergeron:1986} with high H\,I column densities
of $N({\rm H\,I})\approx10^{18}-10^{22}$ \citep[][]{Rao:2006}.
At $z\gtrsim2$, C\,II $\lambda 1334$
absorption can serve as an alternative tracer
of such cool, high column density gas.
Neutral Magnesium and Carbon share similar ionization potential
and we therefore expect the ratio of Mg\,II to C\,II
equivalent width to approximately follow the rest-frame wavelength
ratios, $W_{\rm r}(2796)/W_{\rm r}(1334)=2.09$ for
saturated systems with line-widths dominated by non-thermal
broadening.
Empirically, low-redshift galaxies with constraints on both
Mg\,II and C\,II equivalent widths from \cite{Werk:2013}
exhibit a mean ratio of 
$W_{\rm r}(2796)/W_{\rm r}(1334)=1.7$.
In order to compare the cool gas contents around high
redshift quasars with the sample in this work,
we adopted this mean ratio to convert from
the C\,II equivalent widths reported in \cite{Prochaska:2014}
to expected Mg\,II equivalent widths (consistent with the
conversion from \cite{Prochaska:2014}).
To maintain a uniform cut in absorption
sensitivity, we restricted the quasar sample from
\cite{Prochaska:2014} to those with signal-to-noise
in the background spectrum sufficient to detect a
C\,II $\lambda 1334$ absorption system of
$W_{\rm r}(1334)>0.18$ \AA\, at $3$-$\sigma$
significance. This sensitivity corresponds to the
$W_{\rm r}(2796)>0.3$ \AA\, sensitivity
requirement adopted for the $z\approx1$ samples
with constraints on Mg\,II absorption.

\section{Mg\,II absorption near quasars}
\label{section:absorption}

With the quasar and absorption-line measurements
described in Section \ref{section:sample} in hand,
we characterize the Mg\,II absorption found around
quasars as a function of projected distance, luminosity,
and redshift in Section \ref{section:dependence}.
In addition, we explore the kinematics of the
absorption systems in Section \ref{section:kinematics}
and discuss associated Mg\,II absorption systems found
along the sightline to the foreground quasar itself
in Section \ref{section:associated}.

\begin{figure*}
	\begin{tabular}{cc}
	\includegraphics[scale=0.43]{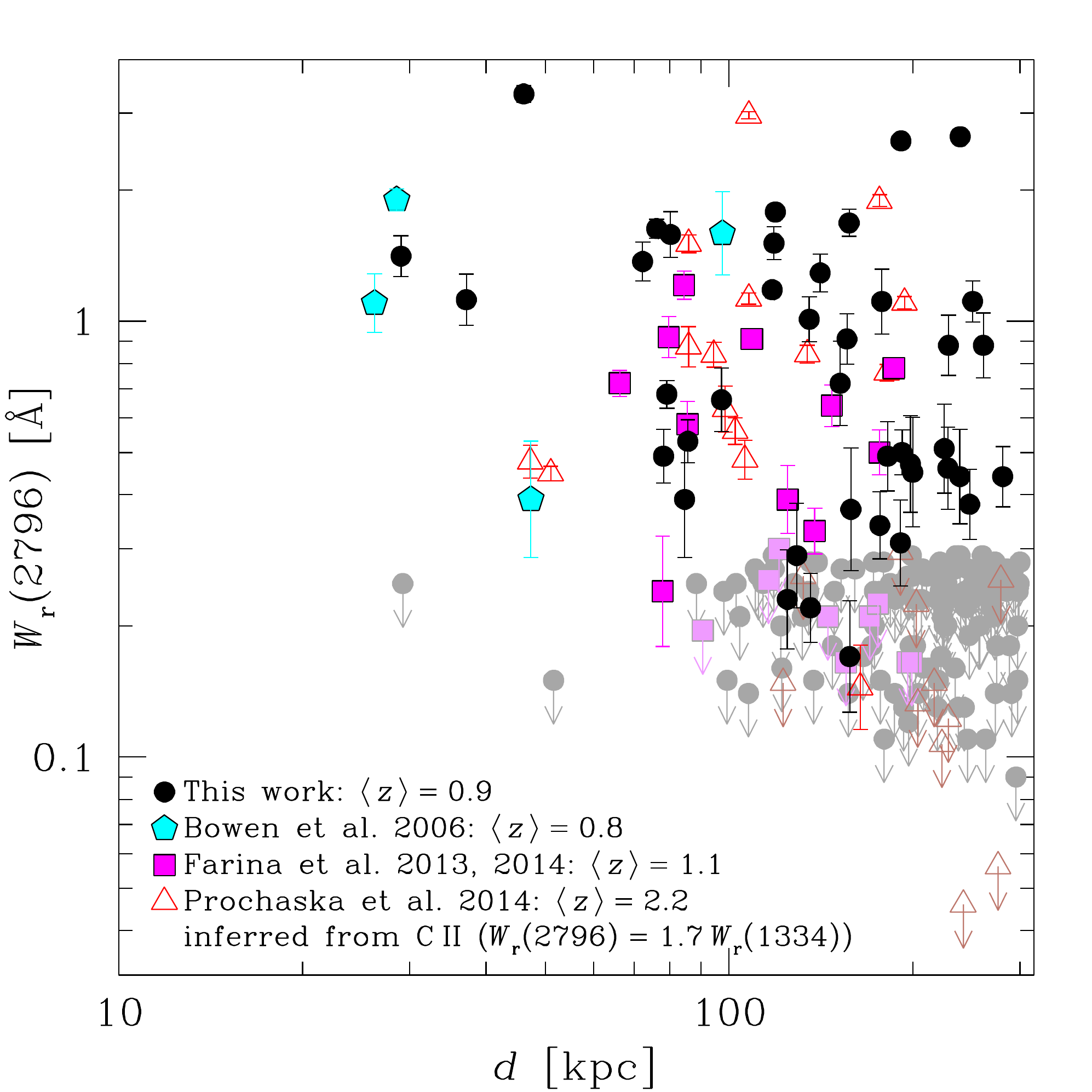} & \includegraphics[scale=0.43]{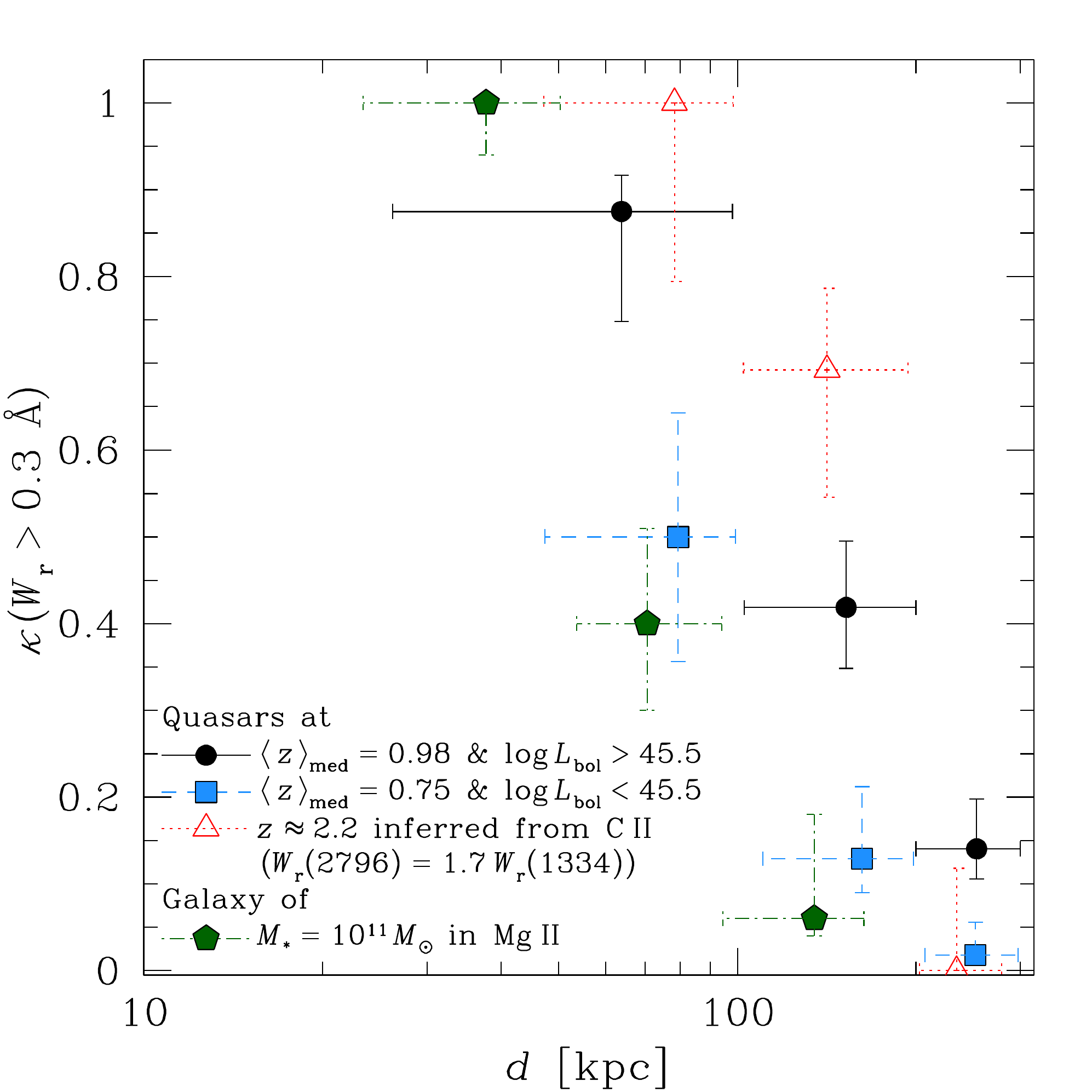} \\
	\end{tabular}
	\caption{{\it left}: Rest-frame Mg\,II $\lambda 2796$ equivalent width versus projected distance.
	              Symbols and coloring are as in Figure \ref{figure:sample}.
	              Non-detections are shown as $3$-$\sigma$ upper limits marked by 
	              downward arrows and lighter shading.
	              {\it right}: Covering fraction for absorption systems with $W_{\rm r}(2796)>0.3$ \AA\,
	              as a function of projected distance for the quasars probed in
	              Mg\,II with luminous quasars ($\log\,L_{\rm bol}/{\rm erg\,s^{-1}}>45.5$)
	              in black solid line and
	              low-luminosity quasars ($\log\,L_{\rm bol}/{\rm erg\,s^{-1}}<45.5$)
	              in blue dashed line.
	              For comparison, the covering fraction for luminous quasars at
	              $z\approx2.2$ based on sample probed in C\,II
	              from \protect\cite{Prochaska:2014} is shown in red dotted line.
	              Finally, we also show the expected covering fraction for
	              galaxies with stellar mass $\log\,M_*/M_\odot=11$ in green
	              dash-dot line.
	              The vertical error bars mark $68\%$ confidence intervals
	              calculated from binomial statistics
	              with a flat prior ($\kappa\in[0, 1]$) and the horizontal error bars
	              mark the full range of projected distances contributing to each bin.
	              The bins are chosen to span the range
	              of $d=0-300$ kpc in $100$ kpc wide intervals.
	              We note that the expected incidence of coincidental Mg\,II absorption
	              systems unrelated to the quasars within the velocity search window
	              is insignificant ($\kappa_{\rm rand}=0.01$) based on the Mg\,II $\partial N/\partial z \partial W_{\rm r}$ measurement
	              from \protect \cite{Zhu:2013}.
	              The SDSS quasars are drawn from a flux-limited survey and consequently, the luminous quasars occur at a higher median redshift of $\langle z \rangle_{\rm med}=0.98$ compared to $\langle z \rangle_{\rm med}=0.75$ for the low-luminosity quasars. The higher gas covering fractions observed for the luminous quasars could therefore be the result of luminosity-dependence, redshift evolution, or a combination of the two. The observed correlation between covering fraction and luminosity is isolated from the possible effects of redshift evolution in Figure \ref{figure:kappa_vs_logLbol_z}.}
	\label{figure:W_kappa_vs_d}
\end{figure*}

\subsection{Dependence on projected distance, luminosity, \& redshift}
\label{section:dependence}
We characterize Mg\,II
equivalent width as a function of projected distance 
in the left panel of Figure \ref{figure:W_kappa_vs_d} and display
the measurements from \cite{Bowen:2006, Farina:2013, Farina:2014},
and \cite{Prochaska:2014} for comparison.
The Mg\,II equivalent widths exhibit an
anti-correlation with projected distance
that is driven by a decrease in incidence at large projected distances.
To verify the significance of this anti-correlation between
equivalent width and projected distance,
we perform a generalized Kendall-$\tau$ rank correlation test
including the non-detections as upper limits \citep{Isobe:1986}.
The Kendall test finds an anti-correlation characterized by a
coefficient of $-0.3$ with a high significance of $p\ll 1\%$\footnote{We note that the measurement uncertainties will tend to
bias the correlation coefficient toward zero and perform this
test to estimate the statistical significance of the anti-correlation
within our sample rather than to measure the coefficient.}.
The anti-correlation is driven primarily by a decrease in the
incidence of absorption systems at $d\gtrsim100$ kpc
rather than by a decrease in the mean equivalent width
of detections.
Interestingly, strong absorption systems of $W_{\rm r}(2796)>1.0$ \AA\,
are observed at $d>100$ kpc, in stark contrast to
$L_*$ galaxies at $z<1.5$ \cite[][]{Chen:2010a, Lovegrove:2011}.

Using a sample of quasars probed in absorption
by background quasar spectra placing constraints on C\,II
absorption, \cite{Prochaska:2014} found excess absorption
for quasars at $z\approx2.2$
compared to quasars
probed in Mg\,II at $z\approx1$
\citep[][]{Farina:2013, Farina:2014}.
 \cite{Prochaska:2014} interpreted this excess as redshift evolution
 in the properties of quasar host halos.
However, the quasars in the $z\approx2.2$ sample from \cite{Prochaska:2014} 
are characterized by high luminosities
(see left panel of Figure \ref{figure:sample}), and it is not clear
that they form a representative sample for the entire quasar
population.
To investigate this possibility, we combine
our SDSS quasars with those reported in the literature
with Mg\,II constraints \citep{Bowen:2006, Farina:2013, Farina:2014}
and divide the resulting combined Mg\,II sample at
the median luminosity into luminous quasars
of $\log\,L_{\rm bol}/{\rm erg\,s^{-1}}>45.5$
and low-luminosity ones of $\log\,L_{\rm bol}/{\rm erg\,s^{-1}}<45.5$.
In doing so, we restrict the combined Mg\,II quasar sample to
those with $z<1.8$ where the sample follows the
well defined redshift-luminosity trend seen in Figure \ref{figure:sample}.
We note that this luminosity division
also corresponds to the traditional
division of AGN into quasars and Seyfert galaxies
\citep[e.g.][]{Schmidt:1983}.

Both the luminous and low-luminosity quasars
exhibit decreasing covering fraction as a function
of projected distance
as shown in the right panel of Figure \ref{figure:W_kappa_vs_d}.
In particular, the low-luminosity
quasar covering fraction decreases from
$\kappa_{\rm Mg\,II}=0.50\pm0.14$ at $d<100$ kpc
to $\kappa_{\rm Mg\,II}=0.13^{+0.08}_{-0.04}$
and $\kappa_{\rm Mg\,II}=0.02^{+0.04}_{-0.01}$
at $d=100-200$ and $200-300$ kpc respectively.
The Mg\,II covering fractions for low-luminosity quasars
are consistent within uncertainties with expectations for inactive galaxies 
with stellar masses $\log\,M_*/M_\odot\approx11$
based on the covering fraction and stellar mass scaling
relation measurements from \cite{Chen:2010a} 
and \cite{Chen:2010b}\footnote{The galaxy sample from  \cite{Chen:2010a} is
based on a sample of $L\approx L_*$ galaxies
at $z\lesssim0.2$, but the scaling relations
in \cite{Chen:2010b}
remain valid at $z<2$ \citep[][]{Lovegrove:2011, Chen:2012, Liang:2014}.}.
A stellar mass of $\log\,M_*/M_\odot\approx11$ is chosen for this
comparison because it is the expected stellar mass of galaxies
with halo masses similar to quasar hosts \citep[][]{Shen:2013}
based on the stellar-to-halo mass relation from \cite{Behroozi:2013}.
On the other hand, the luminous quasars exhibit enhanced
covering fraction that are
a factor of $\approx2$ or more higher than those of the low-luminosity
quasars over the full range of projected distances studied here
with $\kappa_{\rm Mg\,II}=0.88^{+0.04}_{-0.13}$ at $d<100$ kpc,
$\kappa_{\rm Mg\,II}=0.42 \pm 0.07$ at $d=100-200$,
and $\kappa_{\rm Mg\,II}=0.14^{+0.06}_{-0.03}$ at $d=200-300$.

\begin{figure*}
	\begin{tabular}{cc}
	\includegraphics[scale=0.43]{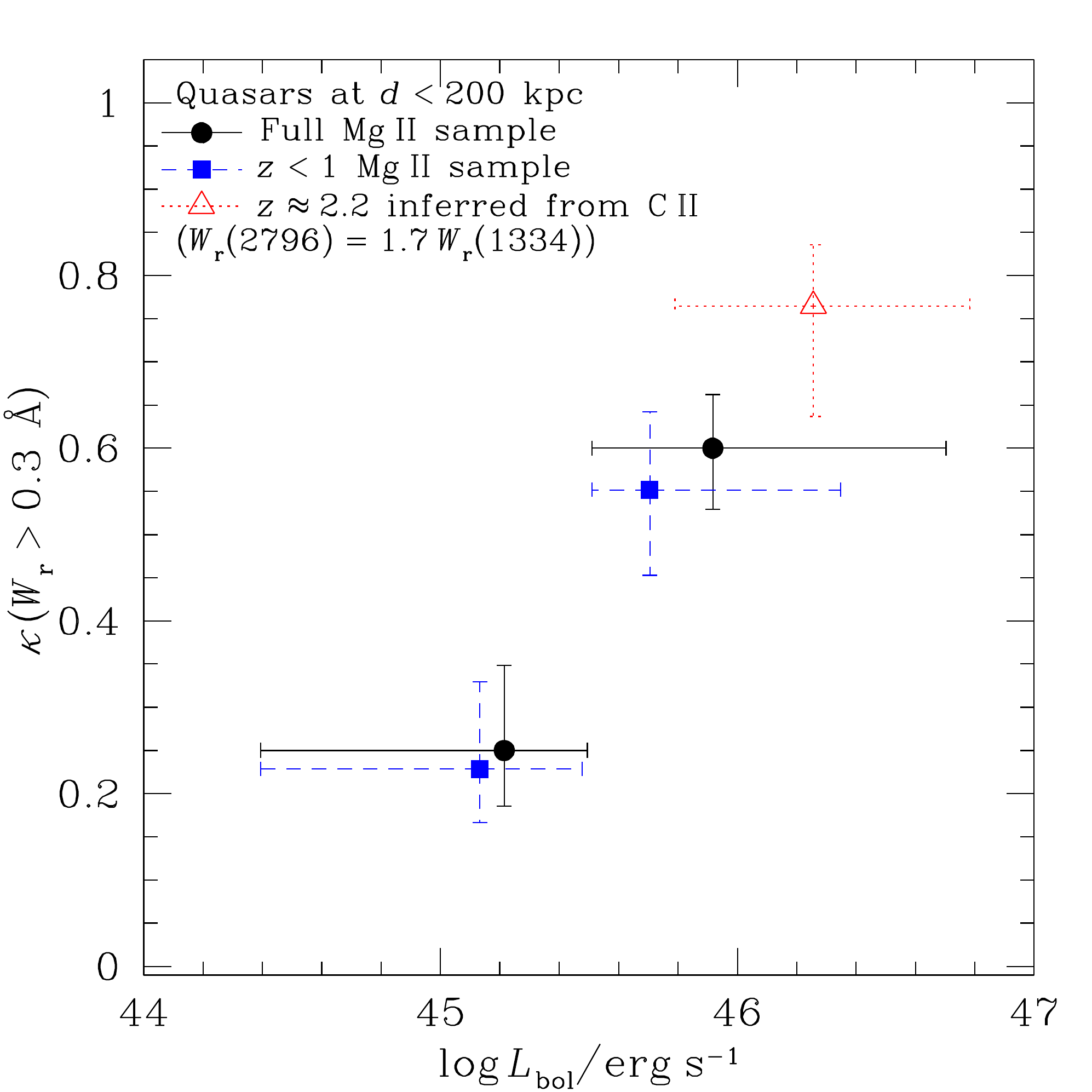} & \includegraphics[scale=0.43]{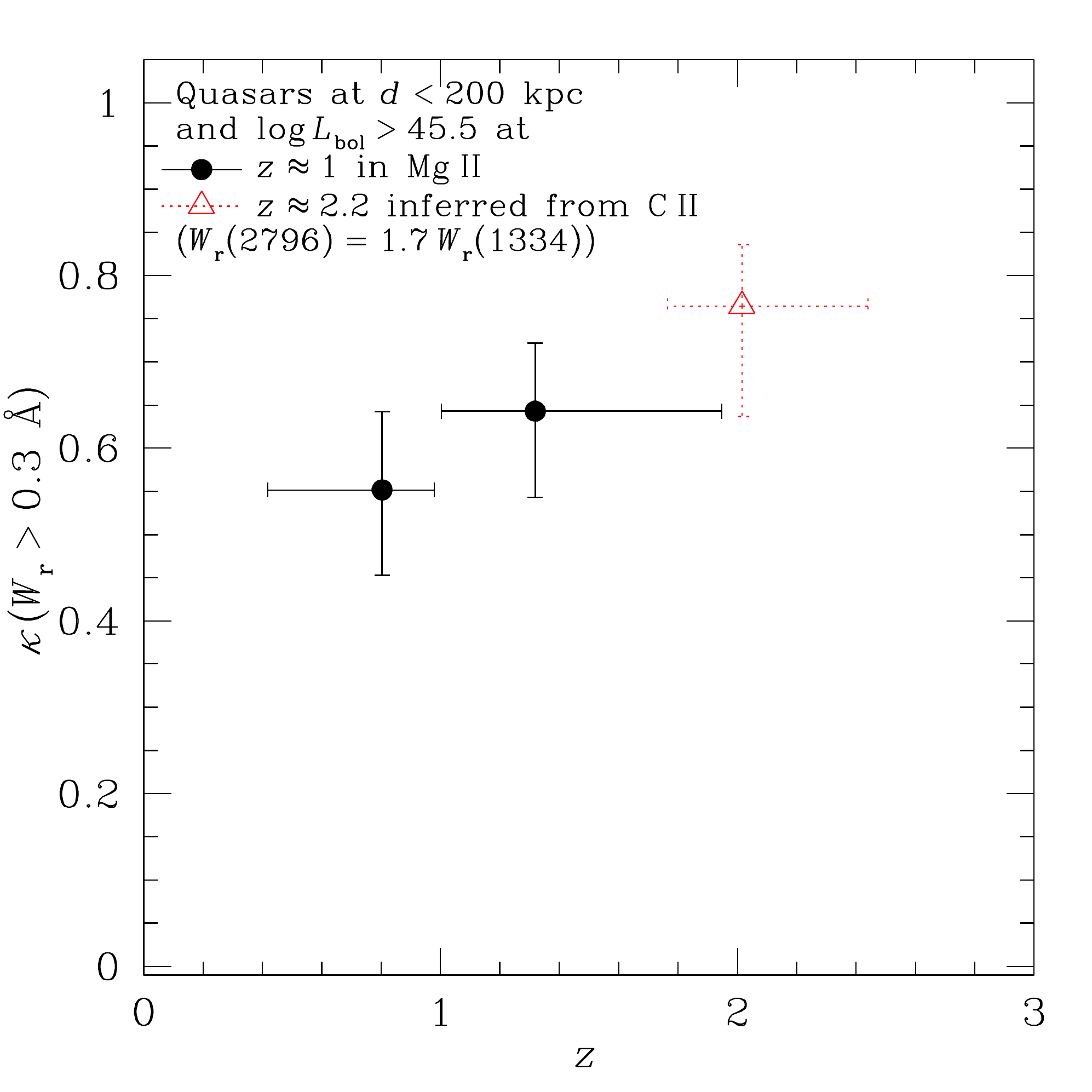}  \\
	\end{tabular}
	\caption{{\it left}: Covering fraction at $d<200$ kpc of
	Mg\,II absorption systems of $W_{\rm r}(2796)>0.3$ \AA\, as a function bolometric luminosity
	              of the foreground quasars. To isolate the correlation with luminosity from
	              possible redshift evolution, we also show the covering fraction as a function
	              of bolometric luminosity with the quasar sample restricted to $z<1$ in blue.
	              This redshift restriction is chosen so that the median redshift of the luminous
	              quasars matches that of the low-luminosity quasars.
	              {\it right}: Covering fraction at $d<200$ for quasars of
	              $\log\,L_{\rm bol}/{\rm erg\,s}^{-1}>45.5$ as a function of redshift.
	              In both panels, the vertical error bars mark $68\%$ confidence intervals
	              estimated from binomial statistics and the horizontal error bars
	              mark the full range contributing to each bin.
	              In both panels, the Mg\,II quasar sample is
	              resampled to mimic the projected distance
	              distribution of the $z\approx2.2$ sample form \protect\cite{Prochaska:2014} (red).}
	\label{figure:kappa_vs_logLbol_z}
\end{figure*}

To verify the significance of the excess Mg\,II
absorption around luminous quasars relative
to low-luminosity ones over the full range
of projected distances studied here,
we also calculate
the covering fraction at $d<300$ kpc
for the low-luminosity quasars
and luminous quasars
and find $\kappa_{\rm Mg\,II}=0.10^{+0.04}_{-0.02}$
and $0.34^{+0.05}_{-0.04}$ respectively.
In addition, we perform
a logrank test \citep[][]{Feigelson:1985}
comparing the distributions of the Mg\,II equivalent
widths for the low-luminosity quasars
and redshift weighted luminous quasars at $d<300$ kpc.
The logrank tests yields $L_{n}=15.2\pm3.7$,
which confirms that the luminous quasars exhibit
excess Mg\,II absorption relative to low-luminosity quasars
with a significance of $p\ll1\%$.

While the low- and high-luminosity quasars exhibit distinct
gas covering fractions, it is possible that the difference in
gas covering fraction is the result of redshift evolution
in quasar host halo gas properties.
In particular, the SDSS quasar sample is drawn from
a flux-limited survey, and consequently the
luminous quasars in our sample occur a higher median redshift
of $\langle z \rangle_{\rm med}=0.98$ compared to
$\langle z \rangle_{\rm med}=0.75$ for the
low-luminosity quasars as shown in Figure \ref{figure:sample}. The excess
absorption observed around the luminous quasars
could therefore be the result of luminosity dependence,
redshift evolution, or a combination of the two.
To isolate the luminosity dependence, we restrict
the luminous quasar sample to $z<1$ in order to
match the median redshift of the luminous
quasar sample to that of the low-luminosity
quasars. The covering fraction for the
luminous quasars at  $z<1$
and $d<300$ kpc is $\kappa_{\rm Mg\,II}=0.32^{+0.07}_{-0.05}$
compared to $\kappa_{\rm Mg\,II}=0.10^{+0.04}_{-0.02}$
for the low-luminosity quasars.
A logrank test comparing the Mg\,II equivalent width
distribution of the low-luminosity quasars with that
of the luminous quasar at $z<1$ confirms this
excess with a significance of $p\ll1$\%.
The comparison of the Mg\,II absorption incidence
between the luminous and low-luminosity quasars
with restricted redshift ranges indicates that the
observed correlation between covering fraction
and luminosity is not being driven by underlying
redshift evolution.

The observed correlation between
Mg\,II absorption covering fraction and quasar
luminosity can explain a significant
portion of the excess cool gas found for quasars at
$z\approx2.2$ relative to those at $z\approx1$.
Nevertheless, the covering fractions for the luminous quasars at $z\approx1$
probed in Mg\,II at $d=100-200$ kpc are somewhat lower
than those of the $z\approx2.2$ quasars probed in
C\,II from \cite{Prochaska:2014} ($\kappa=0.69^{+0.09}_{-0.15}$).
This possible difference between the $z\approx1$ and $z\approx2$
samples could be the result of additional correlation
with luminosity, a difference in projected distance distribution
between the $z\approx1$ and $z\approx2.2$ samples, or
evolution in cool gas content of quasar host halos.

To differentiate these possibilities,
we plot the covering fraction of Mg\,II absorption systems with
$W_{\rm r}(2796)>0.3$ \AA\, at $d<200$ kpc as a function
of luminosity for the combined Mg\,II quasar sample
after resampling to mimic the flatter projected distance distribution
of the \cite{Prochaska:2014} sample
in the left panel of Figure \ref{figure:kappa_vs_logLbol_z}.
The resampling is accomplished by randomly resampling the quasars
probed in Mg\,II without replacement to construct a
maximal possible sample with the projected distance
distribution mimicking that of \cite{Prochaska:2014}.
The random resampling is repeated a large number of
times to measure the mean covering fraction.
Uncertainties in the mean covering fraction are calculated
from binomial statistics with the sample size of the largest
possible subsample with flat projected distance distribution.

The low-luminosity quasars
($\log\,\langle L_{\rm bol}/{\rm erg\,s^{-1}}\rangle=45.2$) exhibit a mean covering
fraction of $\kappa_{\rm Mg\,II}=0.25^{+0.10}_{-0.06}$
while the luminous ones
($\log\,\langle L_{\rm bol}/{\rm erg\,s^{-1}}\rangle=45.9$)
exhibit a covering fraction of 
and $\kappa_{\rm Mg\,II}=0.6^{+0.06}_{-0.07}$
at $d<200$ kpc, indicating a strong
correlation between quasar luminosity and the incidence
of extended Mg\,II absorbing gas.
A similar trend of increasing covering fraction with
luminosity is observed when the sample is restricted
to $z<1$ (see the blue points in Figure \ref{figure:kappa_vs_logLbol_z}).
The trend of increasing covering fraction
with luminosity found for the $z\approx1$ quasars connects
smoothly with the high covering fractions ($\kappa=0.76^{+0.07}_{-0.13}$)
found for high-luminosity quasars
($\log\,\langle L_{\rm bol}/{\rm erg\,s^{-1}}\rangle=46.2$)
at $z\approx2.2$ from \cite{Prochaska:2014}
(see the left panel of Figure \ref{figure:kappa_vs_logLbol_z}).
We note that in calculating the covering fraction for quasars
from \cite{Prochaska:2014}, we restrict the sample to $z<2.5$
where the \cite{Prochaska:2014} sample closely follows
the redshift-luminosity trend seen in the $z\approx1$
samples (see the left panel of Figure \ref{figure:sample}).

With the correlation between Mg\,II absorption
incidence and quasar luminosity in mind, we search
for any additional dependence of Mg\,II absorption
on redshift. To do so, we plot the covering fraction
at $d<200$ kpc of the luminous quasars
as a function of redshift after
resampling to mimic the projected distance
distribution of the \cite{Prochaska:2014} quasar sample
in the right
panel of Figure \ref{figure:kappa_vs_logLbol_z}.
At $z<1$ and $z=1-2$, the luminous quasars
probed in Mg\,II exhibit covering fractions of 
$\kappa_{\rm Mg\,II}=0.55^{+0.09}_{-0.10}$
and $0.64^{+0.08}_{-0.10}$ at $d<200$ kpc,
consistent with no evolution between
$z<1$ and $z=1-2$.
The corresponding covering fraction for the $z\approx2.2$ sample
from \cite{Prochaska:2014} is $\kappa=0.76\pm0.12$,
consistent with the covering fractions found for the
$z<1$ quasars at the level of $1.3\sigma$.
The increase from $\kappa_{\rm Mg\,II}=0.53^{+0.11}_{-0.12}$
at $z<1$ to $\kappa=0.76^{+0.07}_{-0.13}$ at $z\approx2.2$
is suggestive, but
a larger sample of quasars at $z\approx2$ is required to investigate further.
In addition, optical spectra of background sightlines toward quasars at $z<2.5$
can provide constraints on Mg\,II absorption which would eliminate the additional
systematic uncertainty in converting between Mg\,II and C\,II absorption
equivalent widths.

\subsection{Kinematics}
\label{section:kinematics}

\begin{figure*}
\begin{tabular}{cc}
	\includegraphics[scale=0.42]{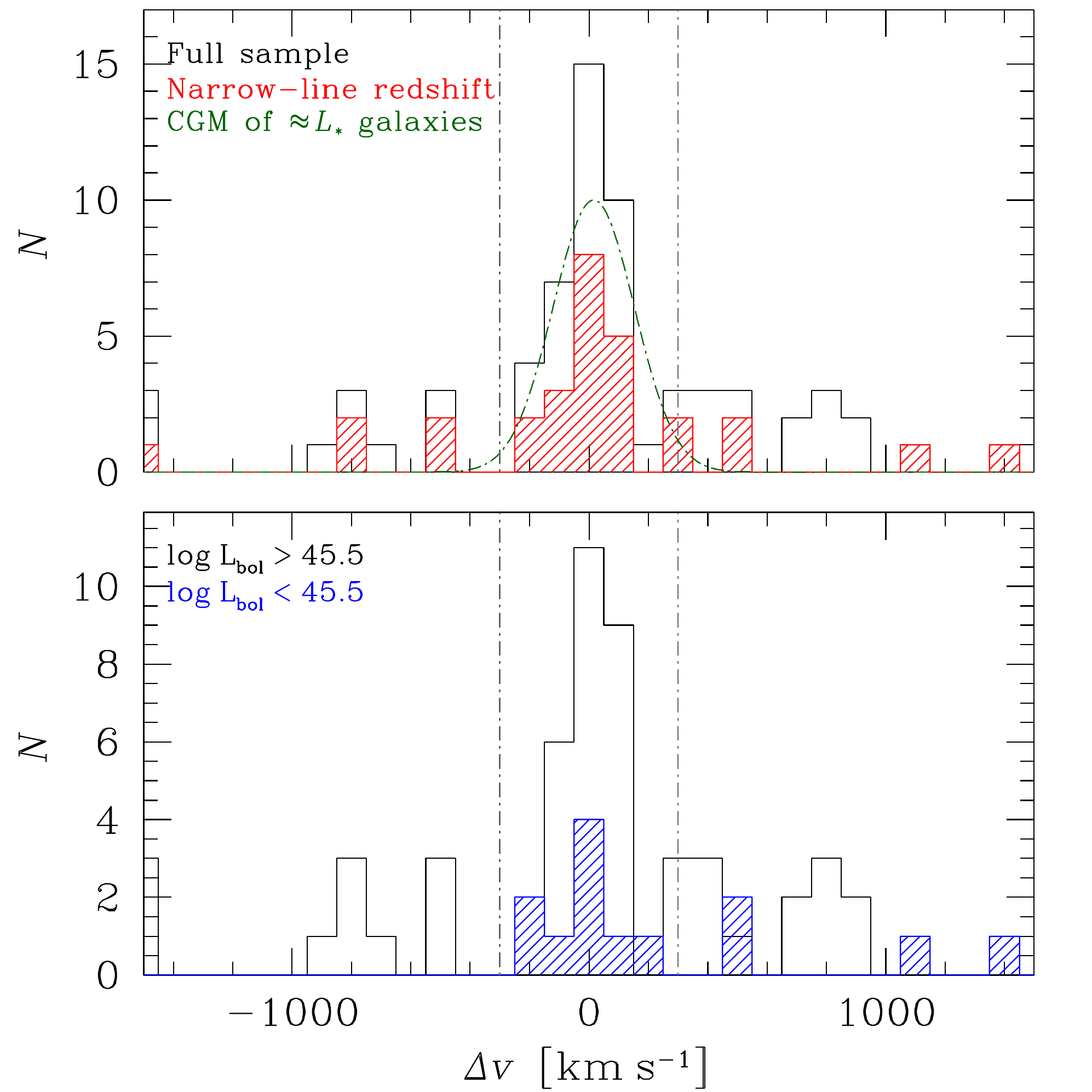} & \includegraphics[scale=0.42]{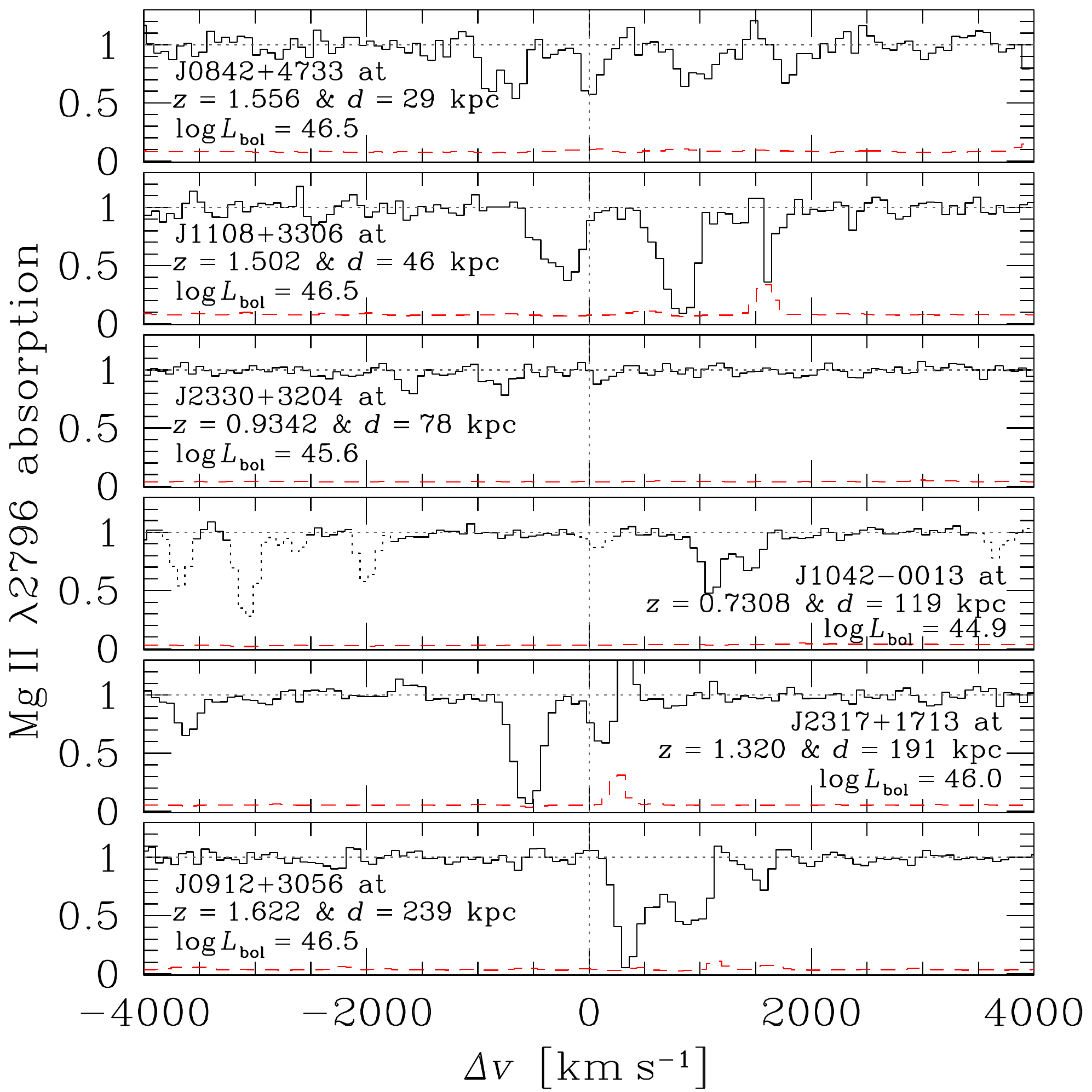}\\
	\end{tabular}
	\caption{{\it left}: Histograms of the radial velocity differences between the quasar systemic redshifts and Mg\,II component redshifts at $\Delta v < 1500$ \kms. The {\it top} panel shows the full Mg\,II sample in black, quasars with narrow-line redshifts in shaded red, and the fit to galaxies from \protect \cite{Chen:2010a} in green dash-dot line. Vertical lines mark $\pm300$ \kms\,, the expected virial velocity of a typical quasar host halo at $z=1.0$. The {\it bottom} panel shows the radial velocity histogram for the high- and low-luminosity quasars in black and blue filled histogram respectively. {\it right}: Kinematic structure seen in absorption toward six
	foreground quasars observed with resolved velocity structure
	in SDSS spectra of the background quasar. The panels
	are labelled with the foreground quasar name, redshift, projected distance,
	and bolometric luminosity. Quasar redshifts from narrow lines are shown to
	four decimal places and those from broad lines are shown to three.
	To prevent confusion between kinematic
	structure and the weaker Mg\,II doublet member, we model the Mg\,II doublet
	observed in the SDSS spectra and plot the data divided by the model for the $\lambda 2803$
	doublet member.}
	\label{figure:components}
\end{figure*}

To characterize the kinematics of the Mg\,II absorption
systems around quasars, we plot histograms
of the radial velocity differences between the quasars
and Mg\,II systems in the left panel of
Figure \ref{figure:components}.
The radial velocity histograms are characterized by
a core that is consistent with the velocity spread
found around galaxies \citep[$\sigma\approx150$ \kms;][]{Chen:2010a}, but
$29$ out of $67$ Mg\,II components ($43\%$) are found
at radial velocity differences with $|\Delta v|>300$ \kms\,
compared to $2$ out of $47$ galaxies ($4\%$)
in \cite{Chen:2010a}.
Ten of $29$ quasars with narrow-line redshifts
and detected Mg\,II absorption
are found to have Mg\,II absorption at $|\Delta v|=300-1500$ \kms\,
($34\%$) indicating that the majority of the
large velocity differences
are not the result of the uncertainties
in broad-line based quasar redshifts
(see the top left panel of Figure \ref{figure:components}).
Such large velocity differences are found
among both high- and low-luminosity quasars,
though the sample of low-luminosity quasars
with Mg\,II absorption is small
(see the bottom left panel of Figure \ref{figure:components}).

In addition to large radial velocity differences,
six quasars in the SDSS sample are found to
have multiple, resolved Mg\,II absorption
complexes spread over hundreds to thousands
of \kms\, in the background quasar spectrum.
The complex kinematics found for these
six quasars are visualized in
the right panel of Figure \ref{figure:components}.
The Mg\,II doublet members are separated
by a wavelength difference that corresponds
to $\Delta v=770$ \kms, creating the potential
for confusion between absorber kinematics
and the weaker doublet member.
To avoid this confusion in visualizing the systems of multiple
Mg\,II absorption complexes, we fit the
Mg\,II absorption using the VPFIT
package and plot the data after dividing by the fit for
the $\lambda 2803$ doublet member.
We note that each absorption ``complex'' in the SDSS spectra
would likely be resolved into multiple components
in high resolution spectra.

\subsection{Narrow, associated absorption systems}
\label{section:associated}
A few percent of optically selected quasars
exhibit narrow, ``associated'' absorption-line
systems ($z_{\rm abs} \approx z_{\rm qso}$)
along the sightline to the quasar itself
\citep[e.g.][]{Weymann:1979, Wild:2008, VandenBerk:2008}.
Associated absorption-line systems
could arise in gas associated with the quasar,
the interstellar medium of the quasar host,
or the interstellar/circum-galactic medium of
neighboring galaxies, but
the ionizing radiation from the quasar
is expected to photoevaporate Mg\,II absorbing
clouds out to distances of $\approx 1$
Mpc along the quasar sightline \citep{Hennawi:2007, Wild:2008}.
A recent comparison between quasars with narrow,
associated Mg\,II absorption and a control sample
of redshift and $i$-band magnitude matched quasars
without associated absorption
found that quasars with associated Mg\,II
absorption exhibit slightly enhanced
dust extinction ($E(B-V)\approx 0.03$)
and a $50-70\%$ enhancement in [O\,II]
emission \citep[][]{Shen:2012}.

A search within our foreground quasar sample
identifies four luminous
and six low-luminosity quasars with associated
($|\Delta v| < 1500$ \kms)
narrow Mg\,II absorption systems
along the foreground quasar sightline.
Two of these four luminous quasars
are found to have Mg\,II absorption systems
detected in the background quasar spectra at
$d=150-300$ kpc, consistent with expectations from the
general high-luminosity quasar sample.
None of the six low-luminosity quasars
with associated absorption systems are found
to have Mg\,II systems in the background quasar
spectra at $d=180-275$ kpc, also consistent
with expectations from the general, low-luminosity
quasar sample.
Though the sample of quasars with associated
Mg\,II absorption and constraints on transverse
Mg\,II absorption from background quasars is small,
these results suggest that the large-scale halo gas
contents of quasars with associated absorption are 
not radically different from those found in the general quasar population.

\section{Discussion}
\label{section:discussion}
Using a sample of $195$ quasars at $z\approx1$
with constraints on Mg\,II absorption from background
quasars at projected distances of $d<300$ kpc from the SDSS, we
characterized the cool gas contents of quasar host
halos as a function of projected distance, quasar luminosity,
and redshift. Our main findings are the following:
\begin{enumerate}
	\item Luminous quasars of $\log\,L_{\rm bol}/{\rm erg\,s}^{-1}>45.5$ exhibit enhanced Mg\,II absorption relative to low-luminosity quasars and inactive galaxies of similar mass at projected distances of $d<300$ kpc.
	\item The absorbing gas near quasars exhibits complex kinematics with $30-40$\% of components found at $|\Delta v|=300-1500$ \kms\, from the quasar systemic redshift. and
	\item The incidence of cool gas absorption around luminous quasars does not evolve strongly with redshift between $z\approx1$ and $z\approx2.2$.
\end{enumerate}

In this section, we discuss the possible origins of
the extended Mg\,II absorbing gas near quasars,
correlation with luminosity, and complex kinematics.
Finally, we briefly consider possible avenues of future research.

Galaxies of $L\approx L_*$ at $z\lesssim0.2$
exhibit high Mg\,II covering fractions out
to a gaseous radius that is observed to scale
with stellar mass according to
$R_{\rm Mg\,II} \propto M_*^{0.28}$
\citep[][]{Chen:2010b} and comparisons with
CGM absorption at higher redshift suggest
that the low-redshift scaling relations
remain valid at $z\lesssim2$ \citep[e.g.][]{Lovegrove:2011, Chen:2012, Liang:2014}.
Since quasars are thought to reside in massive galaxies
of $L \gtrsim L_*$, previous studies \cite[e.g.][]{Farina:2014, Prochaska:2014}
suggested that the high incidence of C\,II and Mg\,II absorption observed at impact parameters
of $d=100-200$ kpc from quasars are the result of mass scaling of the CGM.
In this scenario, the absorption traces the ``normal'' halo gas of inactive galaxies
with masses similar to quasar hosts.
However, the expected Mg\,II
absorption incidence from inactive galaxies with masses
similar to quasar hosts
($\log\,M_*/M_\odot \approx 11$ based on the clustering
measurement from \cite{Shen:2013} and stellar-to-halo
mass relation from \cite{Behroozi:2013})
is significantly lower
than the covering fraction observed around
luminous quasars (see the right panel of Figure \ref{figure:W_kappa_vs_d}).
Moreover, quasar luminosity and host halo mass are only weakly
(if at all) correlated \citep[e.g.][]{Shen:2013} so the observed correlation between
quasar luminosity and Mg\,II absorption cannot
be explained by mass scaling.

Finally, the large velocity differences observed between
the quasar and absorber redshifts
($30-40\%$ of absorption at $|\Delta v|=300-1500$ \kms)
and complex kinematics
seen in some sightlines (see Figure \ref{figure:components})
are inconsistent with gas gravitationally bound to a halo
with the mean mass of quasar hosts \citep[$\log\,M_h/M_\odot=12.8$;][]{Shen:2013}
which have inferred halo virial velocities of $\approx 300$ \kms.
However, clustering measurements constrain the mean mass of quasar hosts, not
the overall host halo mass distribution \citep[see discussion in ][]{White:2012},
and the large fraction of absorption at $|\Delta v|=300-1500$ \kms\,
 could be explained if $\approx30-40$\% of
 quasars reside in halos with larger masses. To evaluate this possibility,
we measure the radial velocity differences observed between
quasars with narrow-line redshifts and Mg\,II absorption systems, correct
for line-of-sight projection by multiplying by $\sqrt 3$,
and find a population averaged velocity
dispersion of $1000$ \kms\, between the quasars and absorbing gas.
The halo mass corresponding to a virial velocity of
$1000$ \kms\, is $\log\,M_h/M_\odot=14.4$.
Such massive halos are three orders-of-magnitude less abundant
than those of $\log\,M_h/M_\odot=12.8$ \citep[][]{Tinker:2008} making it unlikely
that a substantial fraction of quasars reside in such massive systems.
Direct confirmation of this conclusion will require
redshift surveys to characterize the environments of
quasars with constraints on halo gas properties from
background sightlines.

The high Mg\,II gas covering fractions,
correlation with quasar luminosity,
and kinematics can be explained if
the Mg\,II absorbing gas is the result of:
\begin{enumerate}
	\item the CGM of neighboring galaxies
at $\gtrsim 1$ Mpc scales but at projected distances of $d\lesssim 300$ kpc,
	\item feedback from luminous quasars, or 
	\item debris from the galaxy interactions and mergers thought to trigger luminous quasars.
\end{enumerate}
In the following subsections, we discuss each of these possibilities in turn.

\begin{figure*}
	\includegraphics[scale=0.49]{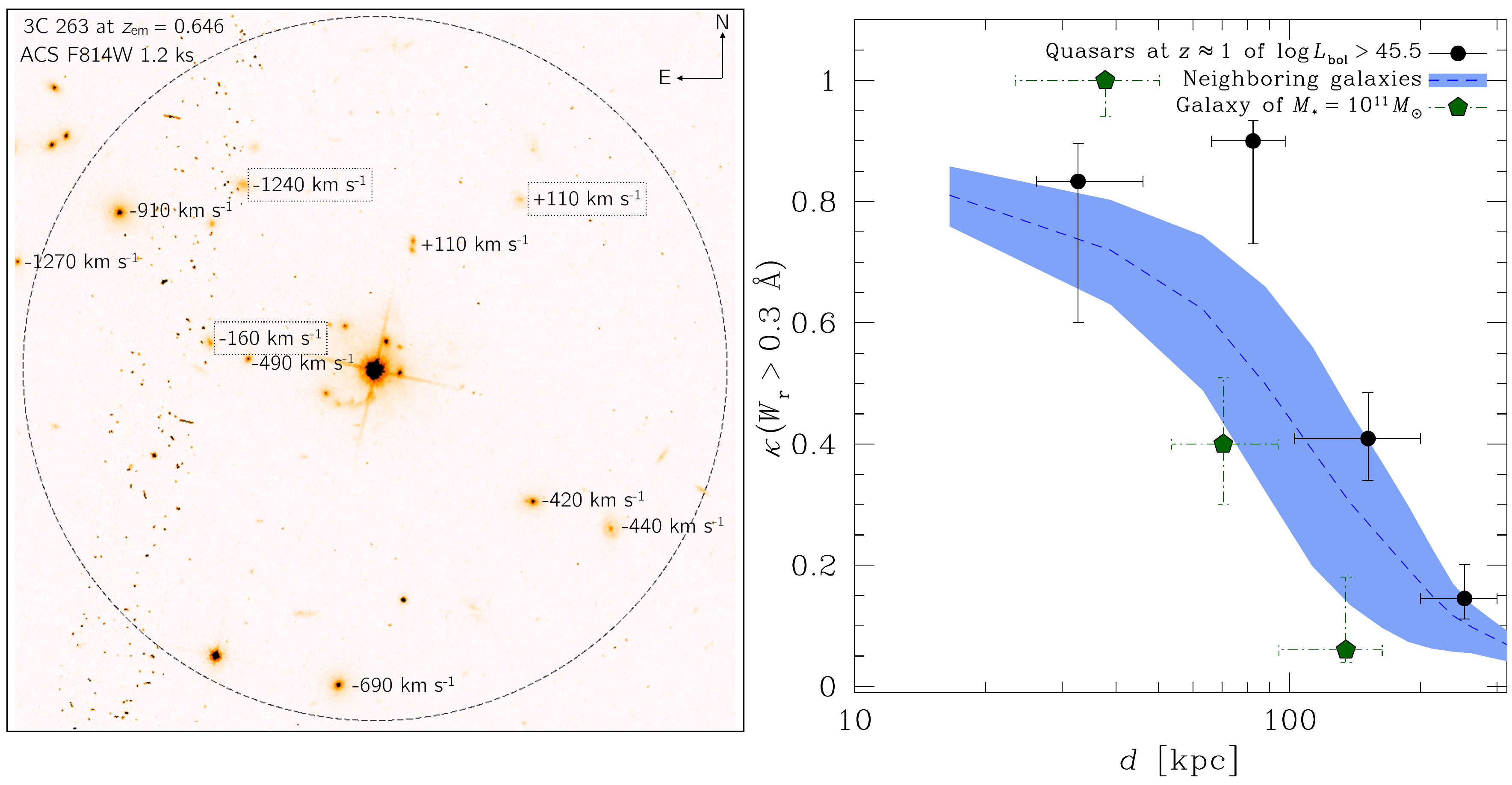}
	\caption{{\it left}: The galaxy environments near a luminous quasar
	              ($3$C\,263 at $z=0.646$)
	              with available deep redshift surveys targeting galaxies of $r_{\rm AB}<23$.
	              Galaxies with secure redshifts and within $|\Delta v|<1500$ \kms\,
	              of the quasar redshift are labelled with $\Delta v$ and star-forming
	              galaxies are highlighted with dotted black outline. 
	              A dashed black circle of $150$ kpc in radius centered on the quasar
	              is shown for scale. The image is a stack of two $600$ second exposures from the
	              Advanced Camera for Surveys on the {\it Hubble Space Telescope}
	              (PI: Mulchaey; PID: 13024) taken with the F$814$W filter with offsets between
	              exposures to fill the gap between the ACS detectors.
	              Cosmic ray removal is not possible in regions
	              with coverage in only one exposure
	              and these regions are visible as $5''$ wide stripes with high cosmic ray
	              contamination.
	              {\it right}: Covering fraction for absorption systems with $W_{\rm r}(2796)>0.3$ \AA\,
	              as a function of projected distance for luminous
	              quasars (black solid line) compared to expectations
	              from a galaxy of $\log M_*/M_\odot = 11$ (green dash-dot line), 
	              and expected incidence due to neighbors of the quasar host (blue dashed line).
	              The mean expected incidence due to neighbors is based on four UV-bright
	              quasars with available deep galaxy redshift surveys and the light blue band
	              represents the estimated uncertainty in the mean.
	              We note, however, that with such a small
	              sample size, uncertainty due to sampling are both large and poorly quantified.}
	\label{figure:deep_images}
\end{figure*}

\subsection{Mg\,II absorption from galaxies neighboring the quasar host}
Large scale quasar-quasar 
\citep[e.g.][]{White:2012} and quasar-galaxy \citep[e.g.][]{Shen:2013}
clustering measurements indicate that quasars
reside in massive halos that trace over-dense
regions of the Universe.
Consequently, the CGM of galaxies in neighboring halos at distances of
$\gtrsim1$ Mpc from the quasar but projected distances of $\lesssim 300$ kpc
could contribute to the Mg\,II absorption
covering fractions and explain the large radial velocity differences
often observed between the quasar and absorber redshifts.

To evaluate the expected covering fraction from
neighboring galaxies, we use
four UV-bright quasars at $z=0.3-0.6$ with highly complete redshift
surveys of galaxies of $r_{\rm AB}<23$ from \cite{Chen:2009, Johnson:2013};
and Johnson et al (in prep).
These surveys reveal the presence of galaxies
with redshifts within $|\Delta v| < 1500$ \kms\, of the
quasar redshift at projected distances of $d\lesssim 300$
kpc that could potentially contribute to the Mg\,II absorption
observed around quasars at similar projected distances
(see left panel of Figure \ref{figure:deep_images}).
We estimate the expected contribution to the covering fraction
observed around quasars from these neighboring galaxies
using the covering fraction and luminosity scaling measurements from
\cite{Chen:2010a}. 
In this calculation, we do not include galaxies more luminous than
$M_B=-22.2$ since such luminous galaxies
exhibit reduced incidence of Mg\,II absorption
compared to galaxies of $L\approx L_*$  \citep[][]{Gauthier:2010}.

The expected Mg\,II covering fractions from neighboring galaxies
can explain a significant portion of the Mg\,II
absorption observed around quasars at $d<300$
kpc (see right panel of Figure \ref{figure:deep_images}).
Moreover, neighboring galaxies are often at
$|\Delta v|\approx300-1500$ \kms\, from the quasar,
potentially explaining the kinematics observed
in Mg\,II absorption.
We note, however, that the sample of quasars with
available deep redshift surveys is small (4 quasars)
and, consequently, the estimated covering fraction
from neighboring galaxies suffers from significant
and poorly quantified sampling uncertainties.

If a substantial portion of the Mg\,II absorption observed around quasars
traces the CGM of neighboring galaxies in other host halos
at Mpc scales, then the correlation
between quasar luminosity and Mg\,II absorption would imply
a correlation between quasar luminosity and quasar-galaxy
clustering.
However, studies of quasar-quasar
and quasar-galaxy clustering found
no correlation between quasar luminosity
and clustering on Mpc scales
\citep[e.g.][]{Croom:2005, Myers:2007, White:2012, Shen:2013}.
In addition, recent observations of projected
quasar-photometric galaxy pair counts find no
evidence for a correlation with quasar luminosity
\citep[][]{Padmanabhan:2009, Zhang:2013, Scott:2015}.
We note that some studies of quasar-photometric galaxy
pair counts find excess counts at $d<300$ kpc
for luminous quasars \citep{Serber:2006}, but this
excess is attributed to galaxies in the quasar host halo
and the excess is not observed on larger scales.

The lack of correlation between quasar luminosity
and clustering are in tension with expectations
from the observed correlation between Mg\,II
covering fraction and quasar luminosity
if a substantial portion of the absorption arises
in neighboring halos on Mpc scales, in apparent contradiction
with the covering fraction estimate shown
in the right panel of Figure \ref{figure:deep_images}.
This discrepancy can be explained by the
proximity effect in which the
UV light from the quasar photo-evaporates
the cool halo gas of galaxies out to Mpc scales.
Indeed, galaxies at $\Delta v<3000$ \kms\, from quasars
exhibit significantly reduced cool halo gas content compared
to the general galaxy population \citep[e.g.][]{Pascarelle:2001}.

\subsection{Mg\,II absorption due to quasar feedback}
Radio-loud, lobe-dominated quasars \citep[class FRII;][]{Fanaroff:1974} are known to drive
powerful outflows
at both high and low redshift
\citep[e.g.][]{Nesvadba:2008, Fu:2009}.
However, only a small fraction quasars
are radio-loud and lobe-dominated \citep[e.g.][]{Urry:1995, Ivezic:2002},
so radio-mode feedback alone cannot be responsible for the observed
correlation between Mg\,II absorption and luminosity.

Recent observations of spatially resolved [O\,III]
emission around luminous, radio-quiet, obscured (Type 2) quasars
revealed the presence of surprisingly
spherical gaseous nebulae that are
both spatially ($\approx 30$ kpc)
and kinematically ($\approx 1000$ \kms) extended
\citep[][]{Greene:2012, Liu:2013a, Liu:2013b}.
Similar structures are observed in [O\,III]
emission around luminous, radio-quiet, unobscured (Type 1)
quasars \citep[][]{Liu:2014} like those studied in this work.
These observations are most naturally explained
as fast outflows with wide opening angles
driven by radio-quiet quasars \citep[also see][]{Zakamska:2014}.

The quasar driven outflows observed in emission on
scales of $\approx 30$ kpc may naturally explain
the high incidence and complex kinematics of Mg\,II absorption around
quasars if the outflows persist to larger distances
but with gas densities that are too low to be observed
in collisionally excited emission lines. An outflow with
velocity $v_{\rm out}=1000$ \kms\, at $15$ kpc
from the host galaxy nucleus could reach distances
of $\approx 100$ ($200$) kpc in $\approx 10^8$ ($2\times 10^8$) years
assuming that the outflow decelerates due to the gravitational
potential of the host halo.
The timescales required to reach $100-200$ kpc from the
host halo are comparable to the high end of quasar
lifetime estimates
\citep[$10^8$ years;][]{Martini:2004}.
Moreover, these fast, quasar driven outflows are observed to be
effective at luminosities of $\log\,L_{\rm bol}/{\rm erg\,s^{-1}}\gtrsim45.5$
\citep[e.g.][]{Veilleux:2013, Zakamska:2014} and could
therefore explain the observed correlation between extended Mg\,II
absorption and quasar luminosity.

In the outflow scenario, the Mg\,II absorbing gas
represents cool clumps entrained in a hotter outflowing
medium \citep[e.g.][]{Costa:2015}.
Indeed, observations of outflows from
both quasars and ultra-luminous infrared galaxies
reveal copious quantities of highly
ionized gas traced by C\,IV and O\,VI absorption
\citep[][]{Arav:2013, Martin:2015}.
However, in the one foreground quasar
with published high resolution background
quasar spectroscopy, cool gas traced by
C\,II absorption is found spread over
$700$ \kms\, but little absorption
from more highly ionized species is found
\citep[][]{Prochaska:2009}.
The low ionization state of the gas in this
system ($N({\rm C\,II})/N({\rm C\,IV})>10$)
is inconsistent with expectations from
cool gas entrained in a quasar driven outflow,
suggesting that quasar driven outflows alone cannot
universally explain the high absorption covering fractions
and complex kinematics observed around quasars.

Alternatively, the gas kinematics and ionization state
can be understood if feedback from luminous quasars
drives the pre-existing halo gas to large velocities, far
in excess of the sound speed.
Such fast bulk motions could lead to an enhanced fraction
of gas with sufficiently high densities to cool and produce
Mg\,II absorbing clumps. In this way, feedback from
the quasar could convert halo gas generally
observed in high ionization states at projected
distances less than the virial radius
\citep[e.g. O\,VI;][]{Chen:2009, Tumlinson:2011, Johnson:2015}
to lower-ionization states observable in Mg\,II absorption.
Feedback from luminous quasars therefore represents
a viable scenario for explaining the cool halo gas properties
around quasar hosts including kinematics,
correlation with luminosity,
and enhanced covering fractions relative to inactive
galaxies of similar masses.

\subsection{Mg\,II absorption from tidal debris}
Luminous quasars are thought to be fueled by gas supplied
during galaxy mergers while the gas required to fuel
less luminous AGN can be supplied by more secular
processes \citep[e.g.][]{Hopkins:2009}.
The high incidence of absorbing gas around quasars
and correlation with luminosity can be explained if the
gas traces debris from the galaxy mergers that
can trigger luminous AGN.
Tidal debris can obtain velocities that exceed
the escape velocity of the host halo \citep[e.g.][]{Toomre:1972},
possibly explaining the kinematics observed in absorption.
The low ionization state 
($N({\rm C\,II})/N({\rm C\,IV})>10$) observed in the
one available quasar with high resolution background
spectroscopy \citep[][]{Prochaska:2009} is consistent with the ionization state
found in Magellanic Stream sightlines with similar
H\,I column densities \citep[][]{Fox:2014}.
Finally, deep $21$-cm observations of the M81/M82
group reveal that relic gas from galaxy interactions
can extend to cover large areas, possibly explaining
the high Mg\,II covering fractions observed out to scales
of $\lesssim 200$ kpc around luminous quasars.
Tidal debris therefore represents a viable possibility
for explaining observed Mg\,II absorption properties
around quasars.
We note, however, that it is not clear that
tidal debris alone can explain the large fraction
($30-40\%$) of Mg\,II absorption components
found at radial velocities of $|\Delta v|>300$ \kms.

\subsection{Future prospects}
\label{section:neighbors}
The high covering fraction of Mg\,II absorption
observed around quasars, correlation
with luminosity, and complex kinematics observed in our SDSS sample
are not consistent with the absorption expected
from inactive galaxies with masses similar
to typical quasar hosts.
In addition, clustering measurements imply that
high- and low-luminosity quasars are hosted by halos
of similar mean masses \citep[][]{Shen:2013} implying that the correlation between
cool gas covering fraction and quasar luminosity
is not the result of simple mass scaling
of the circum-galactic medium.
Together, these observations imply that a substantial
portion of the Mg\,II absorbing gas does not originate in the  ``normal'' halo
gas of the quasar hosts.
The high covering fraction, correlation with luminosity,
and complex kinematics
can be explained if the absorbing gas is the result of:
(1) the CGM of neighboring galaxies on Mpc scales,
(2) feedback from luminous quasars,
or (3) relics from the interactions thought to trigger luminous quasars.
The first of these scenarios is in tension with the lack
of correlation between quasar luminosity and clustering on Mpc scales
\citep[e.g.][]{Serber:2006, Shen:2013}. The remaining two
scenarios make distinct predictions that can
be tested with additional observations.

If a substantial fraction of the Mg\,II absorbing gas is the result
of quasar feedback, then
we expect the cool gas absorption to be
correlated with the presence of extended
outflows observed in [O\,III] emission \citep[e.g.][]{Greene:2012}.

If, on the other hand, the Mg\,II absorbing gas arises in relics from the mergers
that are thought to trigger luminous quasars, then the
absorption would be correlated with the presence of disturbed host
morphologies and nearby tidal remnants.
At $z\approx1$, searching for such interaction signatures
requires the high resolution imaging capabilities of
the {\it Hubble Space Telescope}.

\section*{Acknowledgements}
It is a pleasure to thank Jenny Greene and 
Claude-Andr{\`e} Faucher-Gigu{\`e}re
for lively discussion and insights that helped shape
this paper.
We are grateful to Yun-Hsin Huang for
discussion of the software used for
the absorption measurements and to Cameron Liang
for suggestions that improved the
presentation of data in the figures
throughout the paper.
In addition, we are grateful to the referee, Joe Hennawi,
who provided constructive and substantive
suggestions that helped improve the paper.

SDJ gratefully acknowledges funding from a National
Science Foundation Graduate Research Fellowship
and from the Brinson Foundation.
JSM and HWC acknowledge partial support for this work from grant HST-GO-13024.001-A.

Funding for SDSS-III has been provided by the Alfred P. Sloan Foundation, the Participating Institutions, the National Science Foundation, and the U.S. Department of Energy Office of Science. The SDSS-III web site is http://www.sdss3.org/.
SDSS-III is managed by the Astrophysical Research Consortium for the Participating Institutions of the SDSS-III Collaboration including the University of Arizona, the Brazilian Participation Group, Brookhaven National Laboratory, Carnegie Mellon University, University of Florida, the French Participation Group, the German Participation Group, Harvard University, the Instituto de Astrofisica de Canarias, the Michigan State/Notre Dame/JINA Participation Group, Johns Hopkins University, Lawrence Berkeley National Laboratory, Max Planck Institute for Astrophysics, Max Planck Institute for Extraterrestrial Physics, New Mexico State University, New York University, Ohio State University, Pennsylvania State University, University of Portsmouth, Princeton University, the Spanish Participation Group, University of Tokyo, University of Utah, Vanderbilt University, University of Virginia, University of Washington, and Yale University.

Based in part on observations made with the NASA/ESA Hubble Space Telescope, obtained from the data archive at the Space Telescope Science Institute. STScI is operated by the Association of Universities for Research in Astronomy, Inc. under NASA contract NAS 5-26555.

This research made use of NASA's Astrophysics Data System (ADS)
and the NASA/IPAC Extragalactic Database (NED) which is operated by the Jet Propulsion Laboratory, California Institute of Technology, under contract with the National Aeronautics and Space Administration.

\bibliographystyle{mn} 
\bibliography{biblio}

\bsp \label{lastpage}

\end{document}